\documentclass[12pt]{iopart}

\usepackage{iopams}  
\usepackage{graphicx}

\newcommand{\imag}{\mathrm{i}}
\newcommand{\sumiN}{\sum_{i=1}^{N}}

\renewcommand{\vec}{\mathbf}
\newcommand{\opvec}[1]{\hat{\vec{#1}}}
\newcommand{\sumij}{\sum_{i<j}}

\begin{document}

\title{Quantum Breathing Mode of Trapped Systems in One and Two Dimensions}

\author{Jan Willem Abraham and Michael Bonitz}
\address{Institut f\"ur Theoretische Physik und Astrophysik, Christian-Albrechts-Universit\"at zu Kiel, Leibnizstra\ss{}e 15, D-24098 Kiel, Germany
}
\ead{abraham@theo-physik.uni-kiel.de}
\author{Chris McDonald, Gianfranco Orlando, and Thomas Brabec}
\address{Department of Physics,
University of Ottawa, Ottawa, ON K1N6N5, Canada}

\begin{abstract}
We investigate the quantum breathing mode (monopole oscillation)
of trapped fermionic particles with Coulomb and dipole interaction in one and two dimensions.
This collective oscillation has been shown to reveal detailed information on the 
many-particle state of interacting trapped systems and is thus a sensitive diagnostics for 
a variety of finite systems, including cold atomic and molecular gases in traps and optical lattics, 
electrons in metal clusters and in quantum confined semiconductor structures or nanoplasmas.
An improved sum rule formalism allows us to accurately determine the breathing frequencies
from the ground state of the system, avoiding complicated time-dependent simulations. 
In combination with the Hartree-Fock and the Thomas-Fermi approximations this 
enables us to extend the calculations to large particle numbers $N$ on the order of 
several million. Tracing the breathing frequency to large $N$ as a function of the coupling parameter
of the system reveals a surprising difference of the asymptotic behavior of one-dimensional and 
two-dimensional harmonically trapped Coulomb systems.
\end{abstract}


\maketitle

Trapped systems are of major interest in many fields of research.
Prominent examples are correlated electrons in metal clusters, e.g.  \cite{baletto2005},
confined plasmas  \cite{amiranashvili2003}, ultracold quantum gases in traps
or optical lattices  \cite{giorgini2008, dalfovo99, bloch2005},
electrons in quantum dots \cite{filinov2001, filinov2000, reimann2002, ashoori96, filinov_prb08}
(``artificial atoms''), excitons in bilayers and quantum wells, e.g. \cite{filinov_cpp01,hartmann05,ludwig_njp08,boening_prb11,schleede_cpp12}
trapped ions  \cite{schweigert95},
and colloidal particles  \cite{tatarkova2002}.
Although these systems may
differ in many physical details,
their theoretical descriptions are often similar.
A key property of finite systems in traps are the low-lying collective oscillations since they serve
as a valuable diagnostic tool for the investigation of time-dependent and static features, e.g.
 \cite{giorgini2008, moritz2003, string}.
The importance of these collective modes is comparable to that of spectroscopy in atomic systems.
On the other hand, from the theoretical point of view, the calculation
of normal modes is interesting as the results
can be used to check the quality of models and of nonequilibrium
simulations.
In this work, we concentrate on the
breathing mode -- the (uniform) radial
expansion and contraction of the system.
This collective mode can easily be 
excited in experiments, e.g. by modulation of the confinement frequency or by a rapid 
compression or expansion \cite{moritz2003}.

For classical systems, the breathing mode is well understood
 \cite{schweigert95,henning2008, olivetti2009}.
However, if one incorporates quantum effects,
the description of the mode becomes more complex.
In our recent works \cite{bauch09, bauch10, Abrahamprb},
we reported results from time-dependent simulations
and presented some unique properties of the
quantum breathing mode. Among others, we discovered that the breathing
mode comprises a superposition of at least
two oscillations, which is a pure quantum effect.
While one of the corresponding frequencies, $\omega_\mathrm{cm}$,
has a universal value that equals twice the trap frequency,
the other one, $\omega_\mathrm{rel}$, is strongly
dependent on various system parameters.
The goal of the present work is to determine how
the frequency $\omega_\mathrm{rel}$ is influenced
by, e.\,g., the particle number, the coupling parameter of the system,
the dimension of the system and the type of binary interactions.
These results are crucial for exploiting the above mentioned diagnostic
potential of the breathing mode. 
An analysis for bosons with contact interaction
in a one-dimensional trap that is complementary to ours
is given in Ref.~\cite{schmelcher13}.


Most of the numerical results in our previous
works are based on time-dependent approaches.
The high numerical effort of these simulations poses strong
restrictions on the accessibility of several physical
parameters. In particular, the particle number
usually does not exceed the values $N\approx 20$ in 
one-dimensional (1D) systems \cite{Abrahamprb}
and $N\approx 6$ in two-dimensional (2D) systems \cite{Brabec13}.
To overcome this problem, a simple semi-analytical estimator
for the breathing frequency has been presented in
our recent work~\cite{Brabec13}. As this estimator
is solely based on equilibrium quantities, it enables one to save the
computational effort of the time propagation
in computer simulations, and, additionally, to
use the breathing frequency as an experimental tool
to determine the kinetic, trap and interaction energies
of trapped systems.
In~\cite{Brabec13}, we validated the accuracy of the
estimator, comparing its predictions
with the results from correlated time-dependent
calculations. 
In this work, we make use of it
to describe physical situations for which
no time-dependent calculations are available. We 
systematically study
one- and two-dimensional systems of charged fermionic particles
in the range from small finite
systems to gases with many particles. Such a joint
analysis of the dimensionality and the particle number
has not been performed so far.

To work out how the equilibrium approach is connected to
existing theories for the collective motions of
many-body systems, we further extend the theoretical 
foundations of the breathing mode, providing a systematic
description in terms of time-dependent perturbation theory.
This allows one to conclude that the breathing mode is
characterized by the spectrum of the initial Hamiltonian.
To avoid an exact diagonalization of this Hamiltonian,
we follow the well-known sum rule 
formalism \cite{lipparini89, stringari82, bohigas79, bohigas76} 
to extract an approximation for
the lowest excitation energy from the ground state.
Furthermore, we show how to improve
the sum rule formulas, which turns out to be important,
especially for small systems.
The sum rules allow us to apply ground-state
methods with strongly reduced computational costs.
Hence, we are able to considerably
extend our previous results for fermionic particles
with Coulomb interaction \cite{Abrahamprb,Brabec13}.
The calculations are performed in the framework
of the Hartree-Fock approximation,
for small weakly coupled systems with significant finite-size effects,
and the Thomas-Fermi approximation,
for the transition to large systems.
One of the major results from our analysis
is that the dimension of the system
has a strong influence on the qualitative behavior
of large systems: With growing particle number,
a one-dimensional system behaves more like
an ideal quantum system (see also Ref.~\cite{Abrahamprb}),
while a two-dimensional system, by contrast,
approaches the classical limit.
We show that this unexpected behavior is
indicated by the lowest frequency
of the breathing oscillation
as well as a localization
parameter which relates the average extension
of the system to that of an ideal quantum system.
Discussing how the total energies
scale in non-interacting and strongly coupled (classical)
systems, we further provide an explanation
for the behavior in terms of simple quantities.

We start our presentation with a brief review
of the quantum mechanical description of the trapped system.
Expressing the excitation of collective modes
in terms of time-dependent perturbation theory,
we lay the foundation for the application of the sum rules.
We briefly recapitulate the most important sum rule formulas
and show how their accuracy can be improved. Using this formalism,
we analyze how the breathing frequency depends on the particle number,
the coupling parameter and the dimensionality of the system.

\section{Theory}

\subsection{Time-dependent Schr\"odinger equation}
We briefly recall the physical
setting which has already been described in our previous
works  \cite{bauch09,bauch10,Abrahamprb}.
We aim at the quantum mechanical description
of $N$ identical particles in a $d$-dimensional space.
The time-evolution of the corresponding wave function is
governed by the
time-dependent Schr\"odinger equation
(TDSE)
\begin{equation}
\label{eq:tdse}
\imag \frac{\mathrm{d}}{\mathrm{d}t} | \Psi(t)
 \rangle = \hat H(t) | \Psi(t) \rangle \;,
\end{equation}
where we set $\hbar=1$.
We assume that the Hamiltonian,
\begin{equation}
\hat{H}(t) = \hat{H}_0 + \hat{H}_1(t) \;,
\end{equation}
consists of the stationary part,
\begin{equation}
\hat{H}_0 = \hat{T} + \hat{V} + \hat{W} \;,
\end{equation}
and an additional perturbation term, $\hat{H}_1(t)$.
The explicit form of $\hat{H}_0$ in the spatial
coordinates $\vec{r} \equiv (\vec{r}_1, \dots, \vec{r}_N)$
is given by the kinetic energy
\begin{equation}
T(\vec{r}) = \sum_{i=1}^N 
- \frac{1}{2m} \frac{\partial^2}{\partial \vec{r}_i^2} \;,
\end{equation}
the trap energy
\begin{equation}
V(\vec{r}) = \sum_{i=1}^N \frac{1}{2}m\Omega^2 \vec{r}_i^2 \;,
\end{equation}
and the interaction energy
\begin{equation}
W(\vec{r}) = \sum_{i<j} \frac{K_\alpha}{|\vec{r}_i-\vec{r}_j|^\alpha} \;.
\end{equation}
If it is not further specified,
we assume that the interaction is characterized
by a repulsive power-law potential, $w(r) \propto 1/r^\alpha$,
with the proportionality constant $K_\alpha$. 
Below we will concentrate on the important cases of Coulomb interaction
($\alpha = 1$) and dipole interaction ($\alpha=3$).
After rescaling all lengths, times and energies
in terms of $l_0= \left(1 / (m\Omega) \right)^{1/2}$, $\Omega^{-1}$
and $\Omega$, respectively,
$\hat{H}_0$ takes the form
\begin{equation}
{H}_0(\vec{r}) = \frac{1}{2}\sum_{i=1}^N \left 
\{-\frac{\partial^2}{\partial \vec{r}_i^2} + \vec{r}_i^2
 \right \}
 +\sum_{i<j} \frac{\lambda}{|\vec{r}_i-\vec{r}_j|^\alpha} \;.
\end{equation}
The dimensionless coupling parameter $\lambda>0$
determines the relative strength of the interaction energy  \cite{Abrahamprb}.
It is given by
\begin{equation}
\lambda = \frac{1}{\hbar\Omega} \frac{K_\alpha}{l_0^\alpha} \;,
\end{equation}
with, e.\,g., $K_1 = q^2 / (4\pi\epsilon)$,
for the Coulomb interaction of particles with charge $q$
and the permittivity $\epsilon$,
and $K_3 = C_\mathrm{dd} / (4\pi)$
for the interaction of polarized magnetic or electric
dipoles, with the corresponding proportionality
constant $C_\mathrm{dd}$ \cite{Lahaye2009}.
While, in general, the interaction of two dipoles depends
on their orientation, here we consider only the case 
of parallel dipoles 
\cite{filinov_2012_coll, odell_2008}. In experiments, parallel alignment
perpendicular to the plane of the dipoles is typically
realized with external fields.  


\subsection{Mode excitation}
To formally include a weak mode excitation by an arbitrary
operator $\hat{Q}$, acting only at the time $t=0$, we specify
\begin{equation}
\hat{H}_1(t) = \eta \delta(t) \hat{Q} \;,
\end{equation}
where $\eta$ is a small real parameter.
In the following, we assume that the eigenstates of $\hat{H}_0$ are given by
$\left\{ |0\rangle, |1\rangle, \dots \right\}$ with
the corresponding eigenvalues
$ \left \{ E_0, E_1, \dots \right \}$.
Furthermore, the system is supposed
to be initially in the ground state $|0\rangle$.
With the help of first-order time-dependent perturbation
theory, it can be shown that the time-dependent
expectation value of an operator $\hat{A}$
without explicit time-dependence
takes the form \cite{Abrahamprb}
\begin{equation}
\label{eq:pert_expect}
\langle \hat A  \rangle(t) = 
\sum_{ij} c^*_i c_j \exp \left\{\imag
\left(E_i - E_j \right) t \right\}
\langle i| \hat A| j\rangle 
\end{equation}
with the constant coefficients
\begin{equation}
\label{eq:pert_expans}
c_k = \delta_{k,0} - \imag
\eta \langle k| \hat Q | 0\rangle \;.
\end{equation}
The breathing mode is excited by the monopole
operator
$\hat{Q} = \hat{\vec{r}}^2 = \sum_i \hat{\vec{r}}_i^2$.
For the typical observable $\hat{A}=\hat{\vec{r}}^2$, one can reduce
Eq.~(\ref{eq:pert_expect}) to the expression
\begin{equation}
\label{eq:pert_ev_final}
\langle \hat{\mathbf{r}}^2\rangle(t)
= \langle 0 | \hat{\mathbf{r}}^2|0\rangle 
- {2\eta} \sum_i | \langle 0 | 
\hat{\mathbf{r}}^2|i\rangle|^2 \sin(\omega_{i,0} t) \;,
\end{equation}
where $\omega_{i,0} = E_i - E_0$ are the mode frequencies.
Equation~(\ref{eq:pert_ev_final}) reveals that
the quantum breathing mode is characterized
by a superposition of sinusoids with different
frequencies. For a full characterization
of the breathing mode, one has to calculate
the eigenvalues of $\hat{H}_0$ and the matrix elements
$ \langle 0 | \hat{\mathbf{r}}^2|i\rangle$. 

\subsection{Separation ansatz}
Although the breathing
motion comprises a variety of possible frequencies,
it was shown in recent works \cite{bauch09, bauch10, Abrahamprb}
that the breathing mode is dominated by 
just two frequencies.
One of these has the universal value
$2\,\Omega$. 
This value can be explained by a
formal decoupling of the wave function
into a center-of-mass (CM) and a relative part, 
\begin{equation}
|\Psi (t)\rangle = |\Psi_\mathrm{cm}(t) \rangle \otimes |\Psi_\mathrm{rel}(t) \rangle \;.
\end{equation}
Such a decoupling is induced by the splitting
of the Hamiltonian  \cite{Kim2001},
\begin{equation}
\hat{H}(t) = \hat{H}_\mathrm{cm}(t) + \hat{H}_\mathrm{rel}(t) \;,
\end{equation}
where the contributions read
\begin{eqnarray}
\hat{H}_\mathrm{cm}(t) &= \frac{N}{2} \opvec{P}^2 + \frac{N}{2} \opvec{R}^2 
 + \eta \delta(t) N \opvec{R}^2 \;, \label{eq:hcm}\\
\hat{H}_\mathrm{rel}(t) &= \sumij \Big\{
\frac{1}{2N} \opvec{p}_{ij}^2
+ \frac{1}{2N} \opvec{r}_{ij}^2
+ \hat w(|\opvec{r}_{ij}|) + \eta \delta(t) \frac{1}{N}\opvec{r}_{ij}^2
 \Big\} \;.  \label{eq:hrel}
\end{eqnarray}
Here, we have introduced the CM
and relative contributions, according to
\begin{eqnarray}
\opvec{O} 		&= \frac{1}{N} \sumiN \opvec{o}_i \;, \\
\opvec{o}_{ij} 	&= \opvec{o}_i - \opvec{o}_j \;.
\end{eqnarray}
Since Eq.~(\ref{eq:hcm})
describes a non-interacting $d$-dimensional oscillator problem for the observables
$
\hat{T}_\mathrm{cm} = N \, \opvec{P}^2 / 2$ 
and 
$
\hat{V}_\mathrm{cm} = N  \,\opvec{R}^2 / 2$,
one can conclude that the quantities
$\langle \hat{V} \rangle(t)$ and $\langle \hat{T}\rangle(t)$
always contain oscillations with the frequency $\omega_\mathrm{cm}=2\,\Omega$.
The contributions from the relative system, however, are non-trivial.
Depending on the coupling parameter
$\lambda$, the dominating frequency
obtains the values
$
\sqrt{3} \, \Omega \leq \omega_\mathrm{rel} \leq 2 \, \Omega $,
for Coulomb interaction,
and 
$
2 \, \Omega \leq \omega_\mathrm{rel} \leq \sqrt{5} \, \Omega $, 
for dipole interaction  \cite{bauch09, bauch10, henningphd}.
This frequency corresponds to the first monopole excitation
in the relative system.

\section{Sum rules}
As has been shown, the value of $\omega_\mathrm{rel}$ can be extracted
from the spectrum of $\hat{H}_0$. For more than just a few
particles, however, an exact computation
of the spectrum is impossible. 
Nevertheless, if it is possible to calculate
the ground state of the system, one can make use
of
the quantum mechanical
sum rules to gain some insight into the spectral properties.
In this section, we give a brief review of the sum rules. 
Comprehensive overviews of the theory were
presented for the study of
collective resonances 
in nuclear physics \cite{lipparini89, stringari82, bohigas79, bohigas76}.
While the application
to quantum gases has been subject of many
recent investigations \cite{astrakharchik2005,string,pedri2008,stringari96},
we use the sum rules for the description of
few-particle systems.
We show how the conventional sum rule formulas can be modified
to achieve very accurate results even for such small systems.

\subsection{Calculation of weighted moments}
The weighted moments are defined by 
\begin{equation}
\label{eq:moments}
m_k = \sum_{i=1}^\infty (E_i - E_0)^k \, 
|\langle 0 |\hat{Q}|i\rangle|^2 \;,
\end{equation}
for any operator $\hat Q$ and any integer number
$k \in \mathbb{Z} $.
%
Containing the exact excitation energies,
the moments can be used to define average excitation
energies
\begin{equation}
E_{k,l} = \left(\frac{m_k}{m_{k-l}} \right)^{1/l} \;,
\end{equation}
for positive integer numbers $l$.
In the literature \cite{lipparini89}, one often finds the quantities
$E_{k,2}$
and
$ E_{k,1}$.
The average excitation
energies fulfill the relation
\begin{equation}
\dots \geq E_{k+2,1} \geq E_{k+2,2} \geq E_{k+1,1} \geq E_{k+1,2} \geq \dots
\end{equation}
and, especially,  \cite{bohigas79}
\begin{equation}
\lim_{k \to -\infty} E_{k,1} = \omega_{a,0} \;,
\end{equation}
where the index $a$ corresponds to the lowest state
excited by the operator $\hat{Q}$.
Instead of directly evaluating the sum in Eq.~(\ref{eq:moments}),
one can make use of the sum rules to simplify selected moments.
In the following, we will concentrate on the calculation
of the moments $m_3$, $m_1$
and $m_{-1}$ for the monopole operator
$\hat{Q} = \hat{\vec{r}}^2$.
For $m_1$ and $m_3$, one finds the sum rules
\begin{eqnarray}
m_1 &= \frac{1}{2} \langle 0 | [\hat{Q}, 
	[\hat{H}_0, \hat{Q} ]] |0\rangle \\
	&= 2 \
 \langle 0 | \hat{\mathbf{r}}^2 |0\rangle \;.
\end{eqnarray}
and
\begin{eqnarray}
\label{eq:sumrule_m3}
m_3 &= \frac{1}{2} \langle 0 | 
[ [\hat{Q},\hat{H}_0] , [\hat{H}_0, [\hat{H}_0, \hat{Q} ]]]|0 \rangle \\
&= {8 \langle \hat{T} \rangle + 8 \langle
\hat{V} \rangle + 2 \alpha^2 \langle\hat{W}\rangle} \;.
\end{eqnarray}
The latter result can be obtained, reducing
Eq.~(\ref{eq:sumrule_m3}) to the evaluation of commutators
with the types
$ [\hat{\vec{p}}_i, \hat{\vec{r}}_j]$ and
$ [\hat{\vec{p}}_i, 1 / | \hat{\vec{r}}_j - \hat{\vec{r}}_k|^\alpha ]$
for all occurring indices $i$, $j$, $k$.
The moment $m_{-1}$ can be calculated by
\begin{equation}
\label{eq:inverse_mono}
m_{-1} = 
- \frac{\partial}{\partial \gamma}\langle \hat{\mathbf{r}}^2 \rangle_{\gamma=1} \;,
\end{equation}
where the expectation value refers to the ground state of the Hamiltonian 
\begin{equation}
\hat{H}_0(\gamma) := \hat{T} + \gamma \hat{V} + \hat{W} \;.
\end{equation}
This becomes clear if one writes the derivative
in Eq.~(\ref{eq:inverse_mono})
as
\begin{equation}
\frac{\partial}{\partial \gamma}\langle \hat{\mathbf{r}}^2 \rangle_{\gamma=1} = \lim_{\epsilon \to 0} \frac{\langle\hat{\mathbf{r}}^2 \rangle_{\gamma=1+\epsilon} - \langle\hat{\mathbf{r}}^2 \rangle_{\gamma=1}}{\epsilon}
\end{equation}
and evaluates $\langle\hat{\mathbf{r}}^2 \rangle_{\gamma=1+\epsilon}$
with stationary perturbation theory \cite{bohigas79}.
Henceforth, we will use the notation
$\mathrm{sr}(k,k-l):=E_{k,l}$. 
With the moments $m_3$ and $m_1$, one
obtains the convenient sum rule formula
\begin{eqnarray}
\label{eq:brabecfinalfull}
\mathrm{sr}(3,1)&= \left\{\left( 2+\alpha \right) + 
\left( 2-\alpha \right)
\frac{\langle \hat{T} \rangle}{\langle \hat{V} \rangle} 
  \right\}^{1/2} \nonumber \\
  &= \left\{\left( 2+\alpha \right) + 
\left( 2-\alpha \right)
\left( 1- \frac{\alpha\langle \hat{W} \rangle }{2\langle \hat{V} \rangle} 
\right)
 \right\}^{1/2}\;.
\end{eqnarray}
The two different representations of $\mathrm{sr}(3,1)$ in Eq.~(\ref{eq:brabecfinalfull})
are due to the virial theorem
\begin{equation}
2 \langle \hat{T}\rangle - 2 \langle \hat{V}\rangle + \alpha \langle \hat{W}\rangle = 0 \;.
\end{equation}
For the special case of Coulomb interaction ($\alpha=1$), Eq.~(\ref{eq:brabecfinalfull})
has already been used by Sinha \cite{sinha2000}. A similar result
for contact interaction was derived by Stringari \cite{stringari96}.
Furthermore, we mention that Eq.~(\ref{eq:brabecfinalfull})
is a quantum generalization of the formula by Olivetti \textit{et al.}
for classical systems \cite{olivetti2009}.
A revealing property of
Eq.~(\ref{eq:brabecfinalfull})
is the fact that it allows one to interpret 
the behavior of the breathing frequency with the
ratios of the contributions to the total energy.
Especially the classical limit, with 
$\langle\hat{T}\rangle/\langle\hat{V}\rangle=0$,
and the ideal quantum limit, with
$\langle\hat{W}\rangle/\langle\hat{V}\rangle=0$,
are included. 

Another sum rule formula we will use in this paper
is given by
\begin{equation}
\label{eq:bound_dimensionless}
\mathrm{sr}(1,-1)= \left\{ -2 \frac{\langle \mathbf{r}^2 
 \rangle}{ \left[ \partial \langle \mathbf{r}^2 \rangle/\partial \gamma \right]_{\gamma=1}} \right\}^{1/2} \;.
\end{equation}
It has been presented as a rigorous upper bound of the breathing frequency
by Menotti and Stringari  \cite{string}.

\begin{figure}[tb]
\begin{flushright}
  \includegraphics[scale=1]{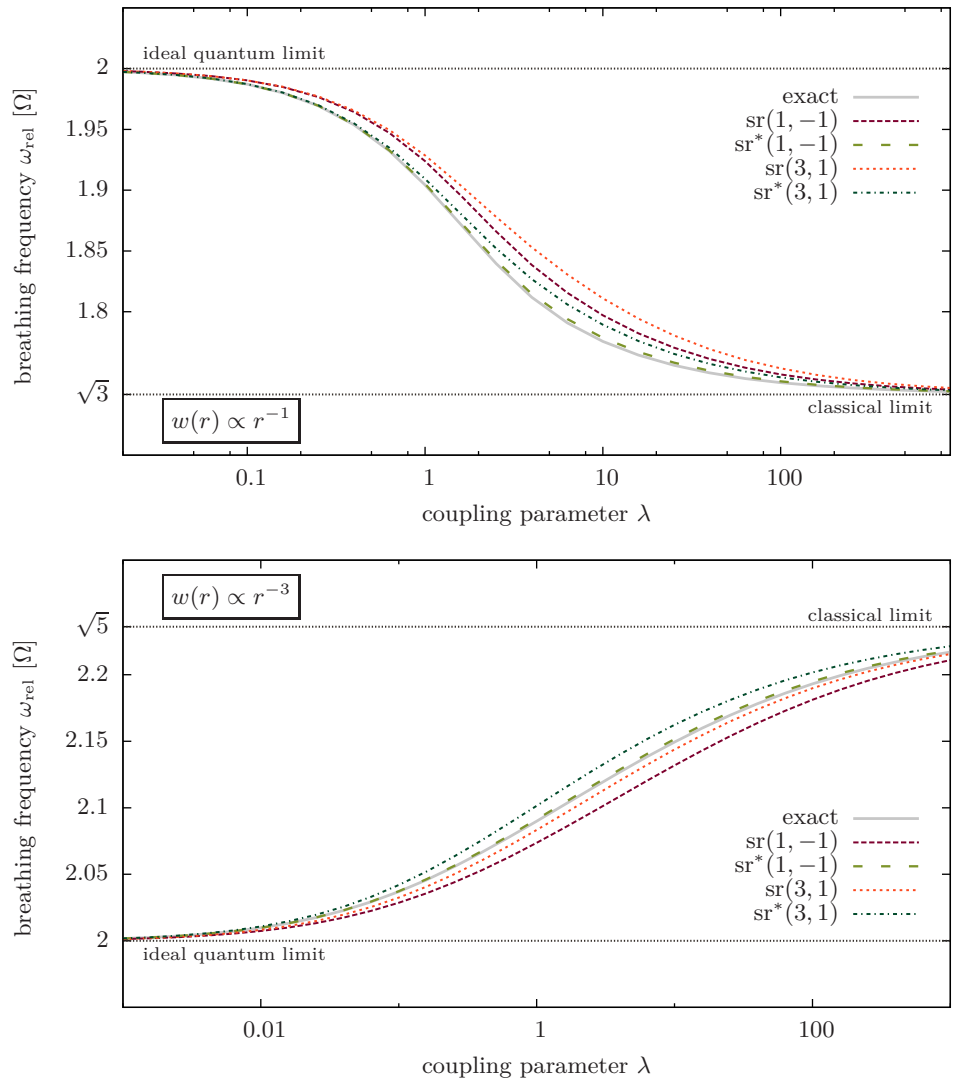} 
\end{flushright}

  \caption{Comparison of the exact breathing frequencies,
  the conventional sum rule formulas and the improved
  sum rule formulas for a spin-polarized
  two-particle system in one dimension
  with Coulomb interaction (top) and dipole interaction (bottom).}
  \label{fig:twopart_sr}
\end{figure}

\subsection{Improved sum rule formulas}
So far, we expressed the moments
with the eigenstates and the eigenvalues
of the full Hamiltonian
$\hat{H} = \hat{H}_\mathrm{rel}+\hat{H}_\mathrm{cm}$. 
This may be disadvantageous 
for the estimation of $\omega_\mathrm{rel}$ if
the weights of the center-of-mass terms
are comparable to those of the relative terms.
With the notations
\begin{equation}
\hat{\mathbf{R}}_\mathrm{cm}^2 := N \hat{\mathbf{R}}^2 
\end{equation}
and
\begin{equation}
\hat{\mathbf{r}}_\mathrm{rel}^2 := \frac{1}{N} \sum_{q<r} \hat{\mathbf{r}}^2_{qr} \;,
\end{equation}
this can be understood more formally, expressing the moments
as
\begin{eqnarray}
\label{eq:momentssplit}
m_l &=\sum_{i=1}^\infty \left( E_{\mathrm{rel},i} - E_{\mathrm{rel},0}  \right)^l
	 |\langle 0_\mathrm{rel} | \hat{\mathbf{r}}_\mathrm{rel}^2  | i_\mathrm{rel}\rangle |^2 \nonumber \\
	 &\;\;\;
	 + \sum_{k=1}^\infty \left( E_{\mathrm{cm},k} - E_{\mathrm{cm},0}  \right)^l
	 |\langle 0_\mathrm{cm} | \hat{\mathbf{R}}_\mathrm{cm}^2  | k_\mathrm{cm}\rangle |^2 \;.
\end{eqnarray}
Using the formulas $\mathrm{sr}(k,j)$,
the contributions from the second sum in Eq.~(\ref{eq:momentssplit})
may cause considerable
inaccuracies
of the estimator for
$\omega_\mathrm{rel}=E_{\mathrm{rel},1} - E_{\mathrm{rel},0}$.
Furthermore, it cannot be guaranteed
that the sum rule formulas are upper bounds
of the frequency $\omega_\mathrm{rel}$ if there
exist non-vanishing contributions 
$E_{\mathrm{cm},k} - E_{\mathrm{cm},0} < \omega_\mathrm{rel}$.
For example, this is the case for dipole interaction,
where $\omega_\mathrm{cm} \leq \omega_\mathrm{rel}$
is valid for all coupling parameters.
A simple solution to this problem is to eliminate
the second sum in Eq.~(\ref{eq:momentssplit}).
As all terms are known analytically,
this can simply be accomplished by
introducing the corrected moments
\begin{equation}
\label{eq:momentscorr}
m^*_k := m_k - 2^{k-1}d \;.
\end{equation}
Using these moments,
we introduce the following improved sum rule formulas,
\begin{equation}
\mathrm{sr}^*(k,k-l) := \left( \frac{m^*_k}{m^*_{k-l}} \right)^{1/l}
\;.
\end{equation}
These formulas are not only expected to yield more accurate results
than the conventional formulas, they also restore
the character of the approximation as an upper bound
for all $\alpha$. The special case $\mathrm{sr}^*(3,1)$
is equivalent to the equilibrium formula presented
in Ref.~\cite{Brabec13}.

To demonstrate the difference between the formulas
$\mathrm{sr}$ and $\mathrm{sr}^*$, we exactly determine the spectrum
of the two-particle system. Expressing
the relative vector $\vec{r}_{1,2}$ in the 
hyperspherical coordinates
$\left(\rho, \phi_1, \ldots, \phi_{d-1} \right)$ \cite{montgomery08},
one can reduce the relative problem to the equation
 \begin{equation}
 \label{eq:final2part}
 \hspace{-1cm}
 \bigg\{-\frac{1}{2}\frac{\mathrm{d^2}}{\mathrm{d}\rho^2} 
 + \frac{\rho^2}{2}
 + \frac{\left(l+(d-2)/2 \right)^2 - 1/4}{2\rho^2} 
 + \frac{\lambda}{2^{\alpha/2}\rho^\alpha} -  E_\mathrm{rel}\bigg\}u_l(\rho) = 0\;.
 \end{equation}
We obtain the spectrum by solving
the corresponding eigenvalue problem in 
the matrix representation that
arises from the expansion of $u_l$ in terms
of FEDVR basis functions (see Sec.~\ref{sec:hf}).
As
the spectrum
of the Hamiltonian $\hat{H}_\mathrm{cm}$
is known analytically,
we can directly reconstruct the moments.
In Fig.~\ref{fig:twopart_sr}, 
we compare
the exact breathing frequencies with the sum rule estimators
for Coulomb and dipole
interaction. 
In both cases, the improvement
of the sum rules leads to a higher accuracy. For dipole
interaction, it is shown that, in fact, the
frequencies from the estimators $\mathrm{sr}(3,1)$ and
$\mathrm{sr}(1,-1)$ are below the exact values.
At the same time, it can be seen that this problem
is solved by the improved formulas.

To conclude, we remark that the subtractive
correction of the moments in
Eq.~(\ref{eq:momentscorr}) does not depend on the particle number.
Therefore, it can be expected that the improvement
of the sum rules is only important for small systems,
where the eigenvalues of $\hat{H}_\mathrm{rel}$
and $\hat{H}_\mathrm{cm}$ are of the same order.
Nevertheless, we stress the difference between the results, because
the influence of the CM subsystem was not
mentioned in some other works, e.\,g., Ref.~\cite{pedri2008}
for the case of dipole interaction.

\section{Numerical methods}
In order to obtain the breathing frequencies
of one- and two-dimensional systems, we calculate
the ground-state energies and apply the sum rule formulas.
We use the Hartree-Fock (HF) and the Thomas-Fermi (TF) approximation
to cover the range from small to large systems.
For comparison, we also show some exact configuration interaction (CI)
and time-dependent Hartree-Fock (TDHF) results, which were
obtained in our previous work \cite{Abrahamprb}.
In the following, some details of the methods
are mentioned.

\subsection{Hartree-Fock approximation}
\label{sec:hf}
The Hartree-Fock approximation \cite{szabo96,bonitzteub} reduces
the $N$-body problem to an effective one-body
problem, where the
interactions are taken into account
on the mean-field level.
The central idea of the HF method is to
assume that the solution of the Schr\"odinger
equation is a single Slater determinant
\begin{equation}
|\Psi\rangle = |\phi_1 \dots \phi_N \rangle \;.
\end{equation}
Requiring that the 
set of single-particle spin orbitals
$|\phi_1\rangle, \dots |\phi_N\rangle$
minimizes the total energy $E=\langle \Psi|\hat{H}_0|\Psi\rangle$,
one can derive the Hartree-Fock equations
\begin{equation}
\hat{F} |\phi_k\rangle = \epsilon | \phi_k\rangle \;,
\end{equation}
where 
\begin{equation}
\hat{F} = \hat{h} + \sum_{i=1}^N \hat{J}_i - \hat{K}_i
\end{equation}
is the Fock operator. For the trapped system, $\hat{h}$
takes the form
\begin{equation}
h(\mathbf{r}) = -\frac{1}{2}\frac{\partial^2}{\partial \vec{r}^2}
+ \frac{1}{2}\vec{r}^2 \;.
\end{equation}
The interaction is incorporated by the operators $\hat{J}_i$
and $\hat{K}_i$, which are defined by their actions
\begin{eqnarray}
\hat{J}_i |\phi_k\rangle &= \lambda \langle \phi_i |\hat{w}|\phi_i\rangle|\phi_k\rangle \;, \\
\hat{K}_i |\phi_k\rangle &= \lambda \langle \phi_i | \hat{w}
|\phi_k\rangle  | \phi_i \rangle \;.
\end{eqnarray}
Expanding the spin orbitals
in terms of a single-particle basis, we transfer
the HF equations to a matrix equation.
For one-dimensional systems, the utilized basis is
a finite-element discrete variable 
representation \cite{rescigno2000, balzerfedvr2010}
(FEDVR).
To avoid divergences, the Coulomb potential is regularized, according
to the standard formula \cite{bauch09,bauch10}
\begin{equation}
w(|r-r'|)=\frac{1}{\left[
(r-r')^2+ \kappa^2 \right]^{1/2}} \;.
\end{equation}
with the screening parameter $\kappa$.
In our recent work \cite{Abrahamprb}, we checked
that converged results are obtained with $\kappa =0.1$.
In two-dimensional systems, by contrast, such a screening is not necessary,
as the basis is formed by the eigenfunctions
of the harmonic oscillator in spherical coordinates.
Our implementation for the calculation of the matrix elements
is based on the code \textit{OpenFCI} by Kvaal \cite{kvaal2008}.

\begin{figure*}[tb]
 \begin{flushright}
  \includegraphics[scale=1]{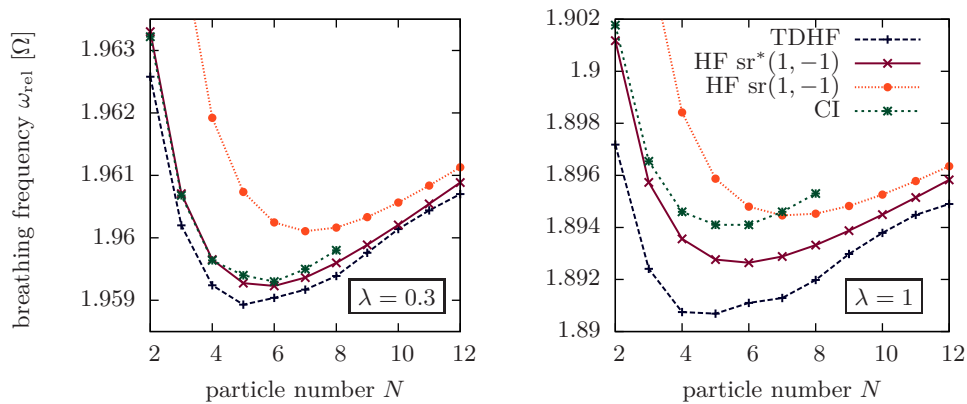} 
  \end{flushright}
  \caption{Breathing frequencies vs. particle number for a  one-dimensional
  spin-polarized system with Coulomb interaction
  and different coupling parameters.
  The results from the improved sum rule formula yield more accurate
  results than the TDHF results
  from our previous work \cite{Abrahamprb}.
  The missing points of the curves $\mathrm{sr}(1,-1)$
  for $N=2$ have the values $1.972\,\Omega$ ($\lambda=0.3$)
  and $1.920\,\Omega$ ($\lambda=1$).}
  \label{fig:sr_1d_omegaN}
\end{figure*}
\subsection{Thomas-Fermi approximation}
For the extension of the HF results,
we use the well-known Thomas-Fermi 
approximation \cite{thomas27,fermi27,spruch91,march57}.
It is expected to yield the correct trend for large particle
numbers, because the oscillations of the single-particle
density -- which
are not shown by the density in TF approximation --
become negligible \cite{Abrahamprb}. 

In the one-dimensional case, the TF equation 
for spin-polarized particles reads
\begin{equation}
\label{eq:tf1d}
\frac{\pi^2}{2} n(r)^2 + \frac{1}{2}r^2 + 
\lambda \int \frac{n(r')}{\left[
(r-r')^2+ \kappa^2 \right]^{1/2}} \mathrm{d}r' = \mu\;,
\end{equation}
where $\mu$ is the chemical potential.
Fixing the particle number $N$ and requiring
the normalization
\begin{equation}
\int n(r) \mathrm{d}r = N \;,
\end{equation}
one has to find the density $n(r)$
and the corresponding chemical potential which
solve Eq.~(\ref{eq:tf1d}).
We obtained our results on a grid,
according to the following scheme:
For each considered particle number,
we start our calculation with a very small coupling parameter
$\lambda$.
At this, we choose the initial trial density
\begin{equation}
n(r) = \frac{1}{\pi} \sqrt{2N-r^2} \;,
\end{equation}
which is exact for a non-interacting system \cite{astrakharchik2011}.
Updating the chemical potential
and the density in a self-consistent procedure \cite{Guelveren2012},
we finally
obtain the density that solves Eq.~(\ref{eq:tf1d}).
After that, we slightly increase $\lambda$
and start a new calculation, where the initial density
is the final density from the previous calculation.
This procedure is repeated until some final value of $\lambda$
is reached.

For two-dimensional systems,
we follow a simpler approach by Sinha \cite{sinha2000}.
Making the ansatz
\begin{equation}
\label{eq:tf2dansatz}
n(\vec{r}) = n(r) = \frac{1}{2\pi \gamma} (r_0^2 - r^2)
\end{equation}
for the density,
we determine the variational parameter $\gamma$
that minimizes the total energy
\begin{equation}
\label{eq:tf_energy}
E = \frac{1}{3}N^{3/2} \frac{1}{\gamma^{1/2}}
+\frac{1}{3}N^{3/2} \gamma^{1/2} 
+ \lambda \frac{512}{315} \frac{\sqrt{2}}{\pi \gamma^{1/4}} N^{7/4} \;.
\end{equation}
In this equation, the single terms correspond
to the kinetic energy, the trap energy
and the interaction energy in this order.

The method for the 2D case has the advantage
that the parabolic ansatz for the
density allows us to express the interaction
energy analytically. In the 1D case,
however, the calculations
are more complex, because the integral
\begin{equation}
\lambda \int \frac{n(r')}{\left[
(r-r')^2+ \kappa^2 \right]^{1/2}} \mathrm{d}r'
\end{equation}
has to be calculated for each grid point $r$
in each iteration step.

\section{Results}
We report the breathing frequencies for one- 
and two-dimensional
systems.
While only spin-polarized systems are treated in
the 1D case,
the occupation of the spin orbitals follows Hund's rules
in 2D systems.
In the following, we consider the finite-size
effects of small systems and the transition
to large systems separately.
In the end of this section, we explain
the calculated breathing frequencies
by the characteristics of the ground states.

\begin{figure}[tb]
 \begin{flushright}
   \includegraphics[scale=1]{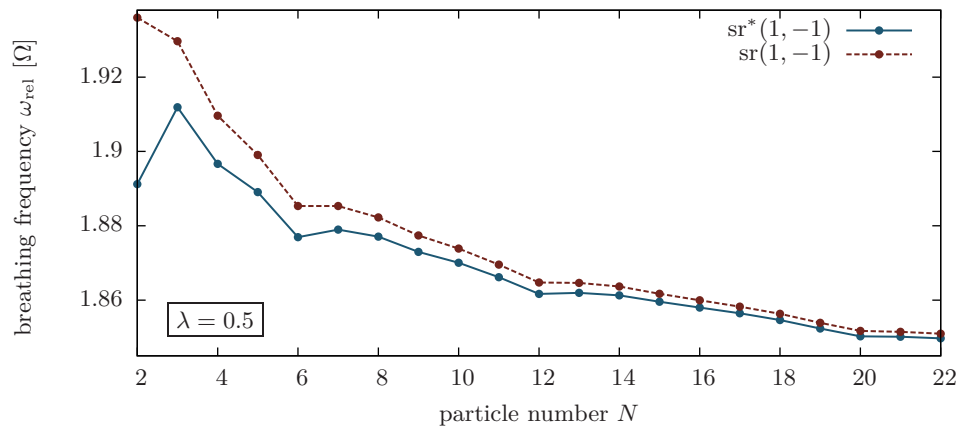} 
  \end{flushright}
  \caption{$N$-dependent breathing frequencies
  for Coulomb-interacting particles in two dimensions.
  The minima correspond to closed energy shells.
  For 2 and 3 particles, the improvement of the sum rule formulas
  is crucial for the correct qualitative description.}
  \label{fig:2d_coulomb}
\end{figure}

\subsection{Small systems}
\subsubsection{1D}$\,$\\
We start with an analysis of small one-dimensional systems.
In Fig.~\ref{fig:sr_1d_omegaN},
we compare the breathing frequencies from different methods
for the coupling parameters $\lambda=0.3$ and $\lambda=1$.
On the one hand, we show the time-dependent HF and CI results
from our previous
work \cite{Abrahamprb}. On the other hand, we
show the results obtained with static HF calculations in combination
with the conventional and the improved sum rule formulas, respectively.
One can draw the following conclusions:
Comparing with the exact CI results, 
the results from the improved sum rule formulas
are more accurate than the TDHF results.
The improved accuracy of the sum rules
can be explained by the fact that the static HF calculations
could be performed with a larger basis than the time-dependent
HF calculations. Furthermore, the error induced by the sum rules
appears to be less important than the error induced by the HF approximation.
Finally, one notices that the improvement of the sum rule formulas
does not only lead to a higher accuracy for small particles, 
it also enables one to find the frequency minimum --
which was already
discovered in our previous work \cite{Abrahamprb} --
at the correct position $N=6$, instead of $N=7$.
However, the figure already reveals that the difference
between both formulas tends to vanish for large particle numbers.

\subsubsection{2D}$\,$\\
The step from 1D to 2D systems is
usually numerically demanding, because 
-- roughly estimated -- the number of
single-particle basis functions has to be squared.
This is especially challenging for time-dependent
calculations.
However, using the sum rules and the HF method, we are
able to investigate the finite-size effects
of the breathing mode in 2D Coulomb systems for the first time.
In Fig.~\ref{fig:2d_coulomb}, we show
the frequencies for small ($N\leq22$) systems
with an intermediate coupling parameter $\lambda=0.5$.
Compared to the one-dimensional systems, we observe
that the non-monotonic behavior of the frequency
has increased. More precisely, the frequency
has local minima for 2, 6, 12 and 20 particles.
These ``magic'' particle numbers are well-known
from various experimental and theoretical
studies of quantum dots \cite{tarucha96,reimann99,reimann2002}.
The corresponding configurations are
very stable, because they are characterized 
by closed energy shells. Typically, experimental
evidence for the occurrence of the magic configurations
is given by the measurement of $N$-dependent addition energies.
Apparently, the breathing mode provides an alternative
tool for the diagnostics of these properties.
Furthermore, it has been shown recently that
important system properties,
such as the kinetic and potential energy, can be obtained
from the results of the breathing frequency \cite{Brabec13}.

Using the HF approximation,
we observe the same non-monotonic behavior
for all
coupling parameters $\lambda \leq 1$.
Another important result is that --
especially for 2 and 3 particles --
the improvement of the sum rules is crucial
to reproduce the correct behavior.

\begin{figure}[tb]
 \begin{flushright}
   \includegraphics[scale=1]{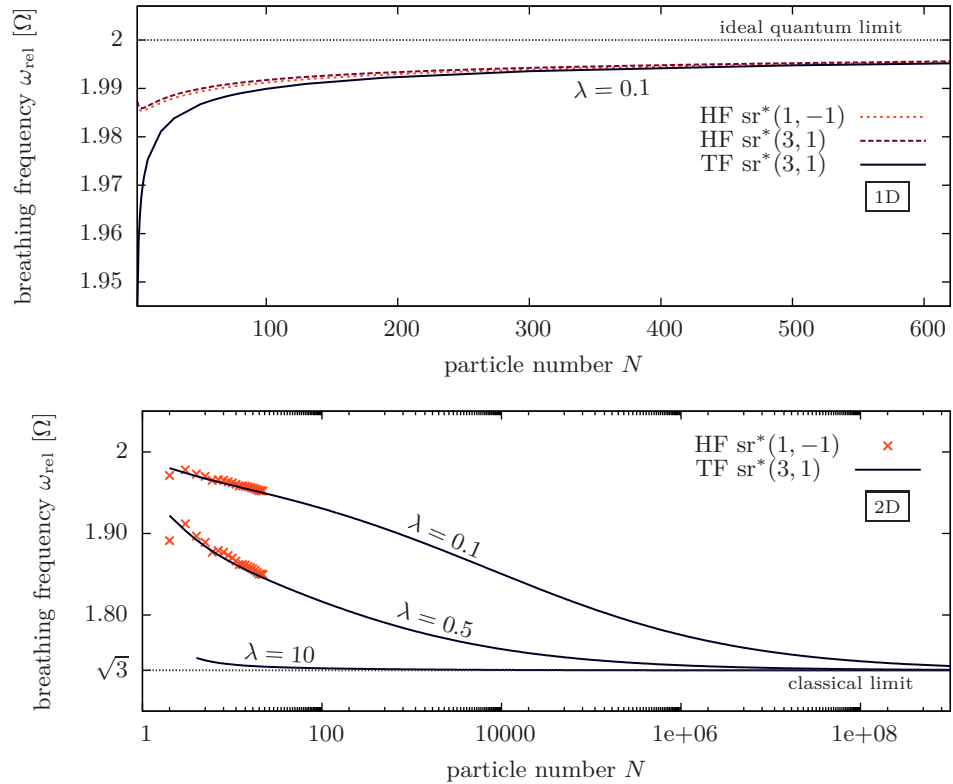} 
  \end{flushright}
  \caption{Behavior of the breathing frequencies for
  large particle numbers in 1D (top) and 2D (bottom).}
  \label{fig:largeN}
\end{figure}

\begin{figure}[tb]
 \begin{center}
   \includegraphics[scale=1.3]{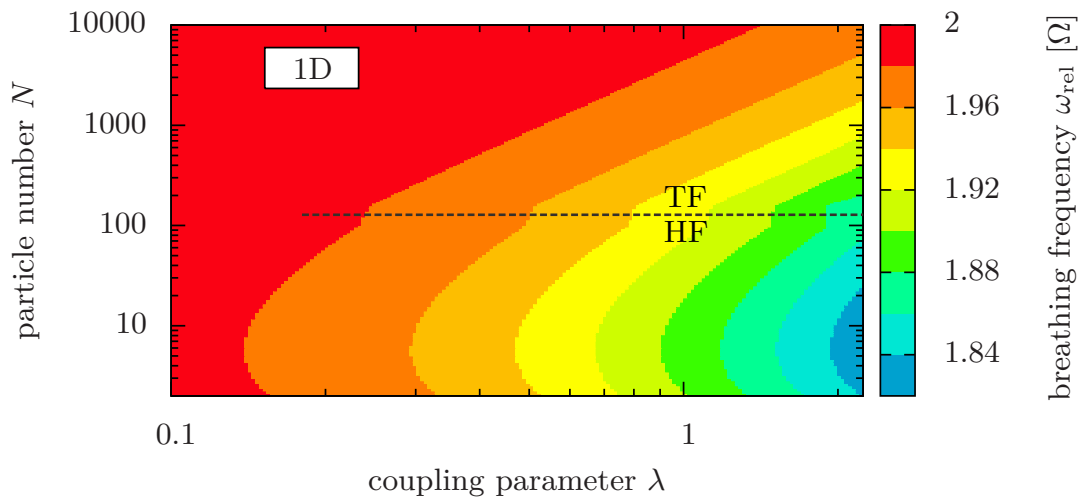} 
  \end{center}
  \caption{
  Breathing frequency of a Coulomb system in a one-dimensional harmonic trap in the $(\lambda,N)$-plane. 
  The data were produced with the Thomas-Fermi approximation
  ($\mathrm{sr}^*(3,1)$, $N > 100$) and the Hartree-Fock approximation
  ($\mathrm{sr}^*(1,-1)$, $N\leq 100$). The step around $N=100$ particles
  is due to the different approximation methods.}
  \label{fig:2dsurvey}
\end{figure}

\subsection{Large systems}
\subsubsection{1D}$\,$\\
In our recent work \cite{Abrahamprb}, we found
several hints that an increase of the particle number
in one-dimensional systems leads to a steady increase
of quantum effects. This means that
the breathing frequency slowly transitions to
the quantum limit $\omega_\mathrm{rel}=2\,\Omega$.
However, the calculations of the frequencies
were restricted to
20 particles with coupling parameters $\lambda \leq 1$.
In this work, we extend the results, reporting the frequencies
for up to 10000 particles with $\lambda \lesssim 2$.

To start the analysis for large systems, we compare
the results from 1D HF calculations and the corresponding
TF calculations for several hundred particles in 
the top plot of Fig.~\ref{fig:largeN}.
As one can see, the difference between both approximations
tends to vanish for large $N$.
Hence, both approximations
confirm the trend $\omega_\mathrm{rel} \to 2 \, \Omega$ 
in the limit $N\to\infty$.
For an overview, the contour plot in
Fig.~\ref{fig:2dsurvey} summarizes
the behavior of 
the breathing frequency in the $(\lambda,N)$-plane.
For all coupling parameters,
the contours have a small step around $N=100$,
where the results from HF and TF calculations
have been joined. Nevertheless, the plot
is suitable to trace the finite size effects,
with the frequency minima for $N=6$, as well as
monotonic behavior of large systems.

As the frequencies approach the ideal limit only very slowly,
we conclude with a remark that
one can give a numerical proof for the large-$N$ behavior
in the TF approximation. For that purpose,
one uses the ideal density to calculate the kinetic energy,
the trap energy
and the interaction energy (with arbitrary
fixed $\lambda$) as a function of the particle number.
Such a calculation reveals that, for each coupling parameter $\lambda$,
there exists a large particle number for which the interaction
energy becomes negligible compared to the other two
contributions. Consequently, the ideal density
becomes a correct solution of Eq.~(\ref{eq:tf1d})
in the limit $N\to\infty$.
In practice, that implies that an increase
of $N$ leads to a faster convergence of the density
if the initial guess in the self-consistent procedure is
the ideal density.

\subsubsection{2D}
\begin{figure}[tb]
 \begin{center}
  \includegraphics[scale=1.3]{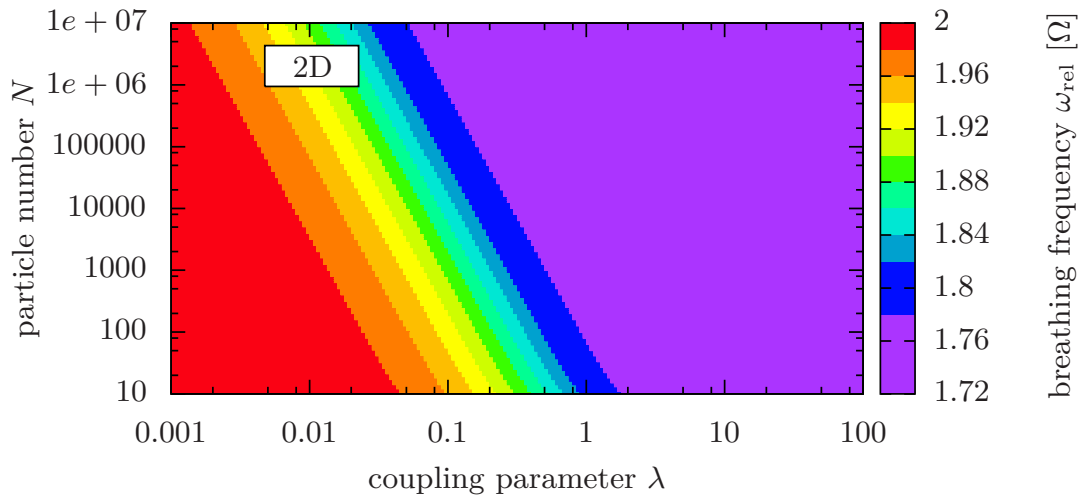} 
  \end{center}
  \caption{
  Breathing frequency of a Coulomb system in a two-dimensional harmonic 
  trap in the $(\lambda,N)$-plane. The data were produced with the Thomas-Fermi approximation
  ($\mathrm{sr}^*(3,1)$).}
  \label{fig:2d_2dsurvey}
\end{figure}

$\,$\\
In two-dimensional systems, the $N$-dependent
behavior of the breathing
frequency is contrary to that of one-dimensional
systems. In the bottom plot of Fig.~\ref{fig:largeN}, one can
see that the frequency always reaches
the value $\sqrt{3}\,\Omega$, in the limit $N\to\infty$.
This behavior is confirmed for a broad range of 
coupling parameters in Fig.~\ref{fig:2d_2dsurvey}.
Hence, 2D systems always approach the classical
limit if either the particle number or the coupling parameter
is increased.

\begin{figure}[tb]
 \begin{flushright}
    \includegraphics[scale=1]{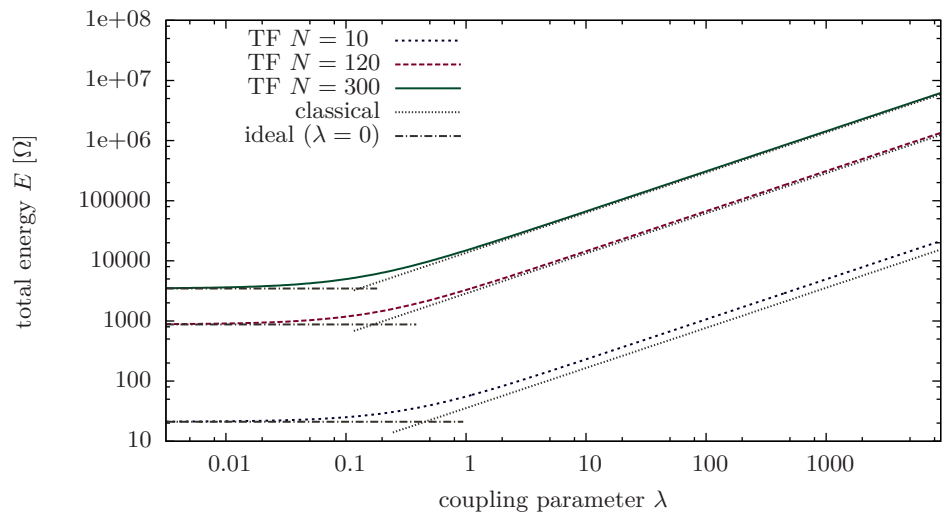} 
  \end{flushright}
  \caption{
  $\lambda$-dependent total energies of a 2D
  Coulomb system for different particle numbers.
  The results from the Thomas-Fermi approximation
  are compared to the results from classical
  molecular dynamics simulations and
  the analytical values of
  an ideal Fermi gas.
  An analogous figure with exact results
  for the 1D case is to be found
  in the work of Astrakharchik and Girardeau \cite{astrakharchik2011}.}
  \label{fig:2d_classical}
\end{figure}

Compared to the results for the 1D systems,
the ranges of numerically accessable $\lambda$ and $N$
are much broader for 2D systems.
This can be explained by the fact that the
parabolic ansatz for the density profile in the Thomas-Fermi
approximation, Eq.~(\ref{eq:tf2dansatz}), is
a drastic simplification of the problem.
Hence, we can treat
all coupling parameters with the same computational
effort of a simple minimization procedure.
Despite the restrictiveness of the parabolic
ansatz, we can show that the results are fairly accurate.
On the one hand,
the good agreement of the TF and the HF results
shown in Fig.~\ref{fig:largeN} justifies the parabolic
ansatz, at least for weakly coupled systems.
On the other hand, 
to check that the ansatz can also be used
to reproduce the correct energies
of strongly coupled systems, we 
show the total energies for a broad range
of coupling parameters and different particle numbers
in Fig.~\ref{fig:2d_classical}.
It turns out that -- for sufficiently
large particle numbers -- the TF results converge
to the results from classical molecular dynamics (MD)
simulations in the limit of large $\lambda$.
%

\subsection{Explanation of the asymptotic behavior
with ground-state properties}
In the following, we provide some supportive
explanation for the different behaviors
of one- and two-dimensional systems.
First, we trace the characteristics
with the help of a localization parameter
that measures the extension of the system.
Second, we introduce a simple estimator
for the distinction between quantum and classical
systems.

\subsubsection{Localization parameter}
\begin{figure*}[tb]
 \begin{flushright}
    \includegraphics[scale=1.3]{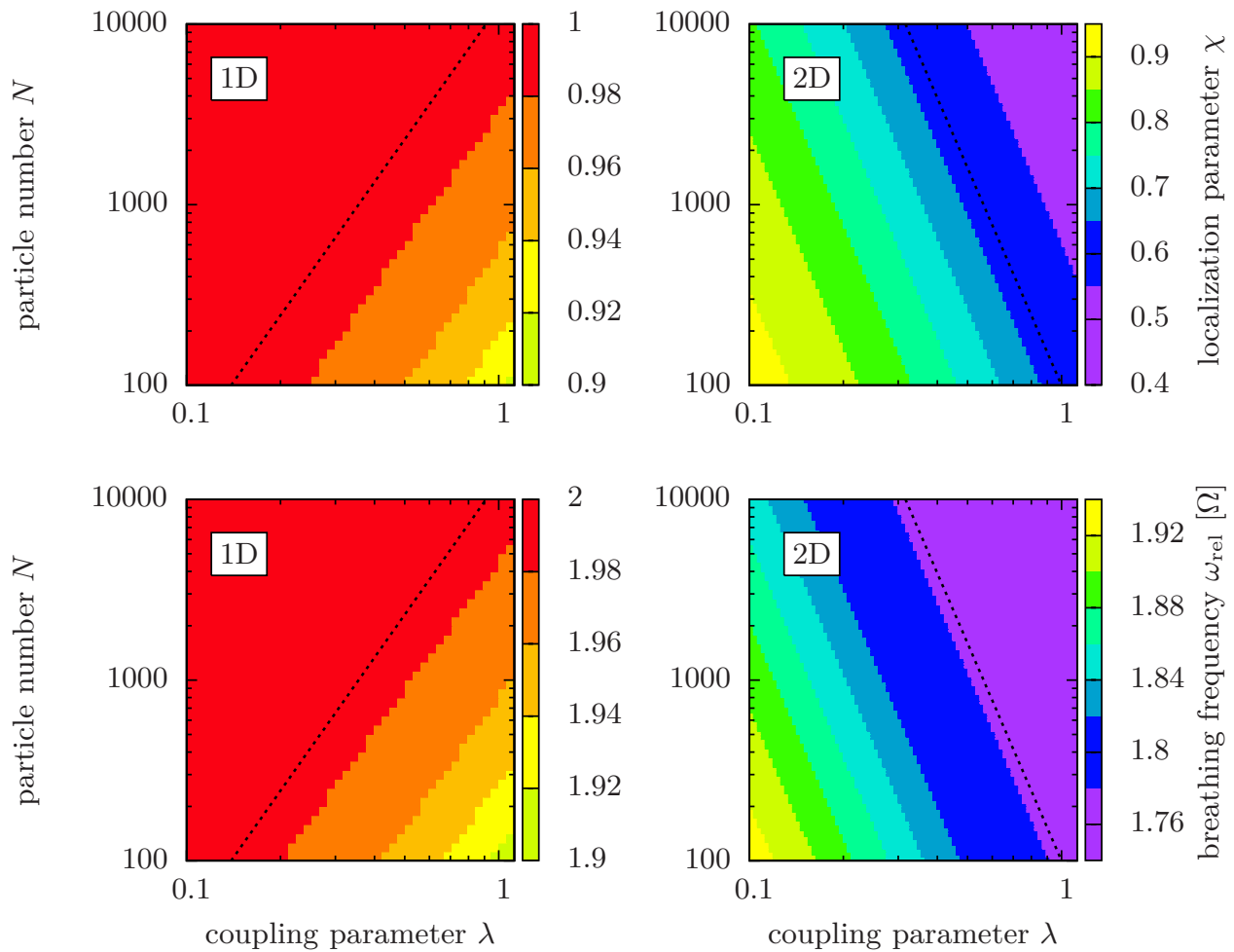}  
  \end{flushright}
  \caption{Comparison of the breathing frequencies (bottom)
  and the localization parameters (top) in the ($\lambda$,$N$)-plane
  for a 1D (left) and a 2D (right) trap. The dotted lines represent exemplary
  constant values.}
  \label{fig:join}
\end{figure*}

$\,$\\
Being inspired by the degeneracy parameter
of a macroscopic homogeneous
electron gas, one can define
a localization parameter for the trap.
This parameter is meant to express the nonideality
of the system with geometric quantities.
An estimator for the mean extension of the system
is given by
\begin{eqnarray}
\sigma
	&= \left \{ 2  \langle \hat{V} \rangle \right \}^{1/2} \;.
\end{eqnarray}
For the non-interaction systems, it has the value
\begin{eqnarray}
\sigma_\mathrm{ideal} &=\frac{1}{\sqrt{2}}
 N \;, &\qquad \mathrm{(1D)} \;,\\
\sigma_\mathrm{ideal} &= \sqrt{2/3} N^{3/4} \;, &\qquad \mathrm{(2D)}  \;.
\end{eqnarray}
With this, one can define the localization parameter
\begin{equation}
\chi = \frac{\sigma_\mathrm{ideal}}{\sigma} \;,
\end{equation}
which measures how much the extension of the 
system deviates from that of an ideal non-interacting quantum system.
Starting with the value $\chi=1$ in the ideal system,
the localization parameter decreases to zero
in the course of the transition
to the strongly coupled regime.
In Fig.~\ref{fig:join}, we compare the localization parameter
with the breathing frequency for one- and two-dimensional systems
in the $(\lambda,N)$-plane.
The data were produced
with the Thomas-Fermi approximation and
the frequencies were estimated with the sum rule formula
$\mathrm{sr}^*(3,1)$.
We concentrate on particle numbers $N\geq 100$
to make sure that finite-size effects are reduced.
Furthermore, since the 1D TF method is reliable
for small coupling parameters, we
restrict the illustration to the regime $\lambda \lesssim 1$.

The figure shows that the localization
parameter is well-suited to track the qualitative
behavior of the breathing frequency.
Equal values of $\chi$ correspond to equal values
of the breathing frequency. 
To illustrate this, the plot shows exemplary dotted lines which indicate
equal values of each quantity.
In the logarithmic plot, these are straight lines.
Remarkably, the lines for 1D systems have a positive slope,
while the lines for 2D systems have a negative slope.
The corresponding best fits in the ranges 
$100\leq N \leq 10000$ and $0.1\leq\lambda \leq 1$ read
\begin{eqnarray}
\label{eq:geradengleichung1d}
N &= (1.3 \times 10^4) \lambda^{2.45} \;, &\qquad \mathrm{(1D)} \;, \\
\label{eq:geradengleichung2d}
N &= 100 \lambda^{-4} \;, &\qquad \mathrm{(2D)} \;.
\end{eqnarray}

\subsubsection{Estimator for intermediate couplings}$\,$\\
Finally, we provide another rough estimator for a supportive
understanding of the observed behavior.
For the two-dimensional
system, the straight lines in Fig.~\ref{fig:2d_classical}
demonstrate
that the total energies can be approximated by those
from non-interacting Fermi gases for weak couplings,
\begin{equation}
E_\mathrm{ideal}^\mathrm{2D}= \frac{2}{3} N^{3/2} \;,
\end{equation}
and those from purely classical systems for
strong couplings, 
\begin{equation}
E_\mathrm{classical}^\mathrm{2D}= K \lambda^{2/3} N^{5/3} \;.
\end{equation}
The last equation is derived by setting the kinetic energy
to zero and minimizing the remaining
terms in Eq.~(\ref{eq:tf_energy}).
The constant $K$ has the value
\begin{equation}
K=  \left( \frac{256\sqrt{2}}{315\pi} \right)^{2/3} 
+ \frac{512\sqrt{2}}{315\pi}  \left( 
\frac{256\sqrt{2}}{105\pi} 
 \right)^{-1/3}  \;.
\end{equation}
As has been shown by Astrakharchik and
Girardeau \cite{astrakharchik2011}, one
can also derive the corresponding terms
for one-dimensional systems,
where one obtains
\begin{equation}
E_\mathrm{ideal}^\mathrm{1D}= \frac{1}{2} N^2
\end{equation}
for non-interacting systems
and
\begin{equation}
E_\mathrm{classical}^\mathrm{1D}= \frac{3}{10} (3 \lambda N \ln N)^{2/3}N
\end{equation}
for classical systems.
In both 1D and 2D, 
we use the quantities
$E_\mathrm{ideal}$ and $E_\mathrm{classical}$
to define a regime of intermediate coupling.
Taking account of the $\lambda$-dependence
of the total energy shown in Fig.~\ref{fig:2d_classical},
we mark this region by the coupling parameter $\tilde{\lambda}$
for which the classical estimator $E_\mathrm{classical}$
is equal to the ideal estimator
$E_\mathrm{ideal}$.
One obtains
\begin{equation}
\label{eq:lambdaIntersect2D}
\tilde{\lambda} = \left( \frac{2}{3K} \right)^{3/2} N^{-1/4}
\end{equation}
for 2D systems,
which is in agreement
with the functional form
of Eq.~(\ref{eq:geradengleichung2d}).
For the 1D systems, one obtains
\begin{equation}
\label{eq:lambdaIntersect1D}
\tilde{\lambda} =\frac{1}{3} \left( \frac{5}{3} \right)^{3/2}
\frac{N^{1/2}}{\ln N} \;.
\end{equation}
In this equation, the relation between
$N$ and $\tilde\lambda$ slightly
differs from that in Eq.~(\ref{eq:geradengleichung1d}).
However, we expect that both definitions
come to agreement if one covers a larger area
of the ($\lambda,N$)-plane shown in Fig.~\ref{fig:join}.

\begin{figure}[tb]
 \begin{flushright}
    \includegraphics[scale=1]{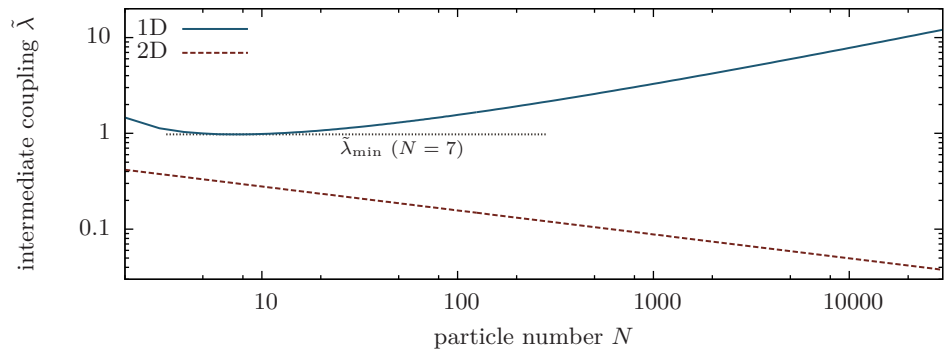} 
  \end{flushright}
  \caption{$N$-dependent behavior of the intermediate coupling
  parameter $\tilde \lambda$,
  for one- and two-dimensional systems.}
  \label{fig:intersect}
\end{figure}

The coupling parameter  $\tilde{\lambda}$
roughly divides the systems into the ones with dominating
quantum-like behavior ($\lambda < \tilde{\lambda}$)
and the other ones with dominating
classical behavior ($\lambda > \tilde{\lambda}$).
Apparently, the trend for large systems
is as follows: An increase of $N$ leads to a growing
classical regime in 2D and a growing quantum regime
in 1D.
In Fig.~\ref{fig:intersect}, Eqs.~(\ref{eq:lambdaIntersect2D})
and (\ref{eq:lambdaIntersect1D}) are plotted.
Remarkably, for one-dimensional systems,
$\tilde{\lambda}$ has a minimum for $N=7$
and thus even reproduces the non-monotonic
behavior.

\section{Discussion}
In this work, we presented an improvement
of the conventional sum rule formulas
for the calculation of the monopole
excitation spectrum. The main idea was the elimination
of the analytically
known center-of-mass contributions to the weighted moments.
It turned out that the improvement yields very accurate
results for the frequencies of the quantum
breathing mode.

In our previous work, we discussed that
the breathing mode is an indicator for the
nonideality of the system  \cite{Abrahamprb}.
With Eq.~(\ref{eq:brabecfinalfull}), we can
express this statement more precisely,
mapping the breathing frequency to a fixed relation
between the kinetic energy, the trap energy
and the interaction energy.
Having analyzed the breathing mode for various
configurations, we can summarize the characteristics
as follows.
If the particle number of the system is fixed
and the coupling parameter $\lambda$ is increased, the frequency
always reaches the classical limit $\sqrt{3}\,\Omega$.
This behavior corresponds to the well-known
Wigner crystallization \cite{wigner34,filinov2001}.
The $N$-dependence of the breathing frequency
is more complicated. {\em One-dimensional
systems} are characterized by a frequency minimum 
for $N=6$, followed by a monotonic increase
of the frequencies until the ideal limit $2 \,\Omega$
is reached. In {\em two-dimensional systems},
the breathing frequency of small systems
reflects the shell structure of the configuration.
For weakly interacting systems, we could observe minimum
frequencies for configurations with closed energy shells.
We expect that this $N$-dependent behavior
will be replaced by the characteristics
of strongly coupled clusters, for large coupling parameters.
The most remarkable difference from one-dimensional
systems is that the frequency reaches the classical
value in the limit $N\to\infty$ with fixed coupling parameter $\lambda$.

To summarize, we state that the sole knowledge of
the parameter $\lambda$ is not sufficient to characterize
the state of an interacting trapped quantum system, in particular, to decide
whether the breathing mode
is dominated by quantum effects or classical effects.
Instead, one also has to take into account the particle number
and the dimensionality of the system.
We provided several estimators for the behavior
of large one- and two-dimensional systems.
The influence of the explicit form of the pairwise interactions
on this behavior remains to be analyzed in future works.
Furthermore, the presented improved sum rules should allow to investigate in more detail
the influence of the spin statistics and
the effects of strong correlations \cite{schmelcher13} 
e.\,g., with path integral Monte Carlo methods \cite{filinov2010, schoof2011}.

We expect that our results will be of interest for a variety of quantum many-body systems confined in harmonic potential wells. 
As an example, we mention ultracold ions and neutral atoms or combinations  of them \cite{peotta_13} in traps. In these systems 
collective modes, including the breathing mode are easily excited \cite{peotta_13}. In Ref. \cite{harlander_11} it was suggested to use 
the sloshing mode of a trapped small ion ensemble to transmit information to a second ensemble. A similar concept might also be possible by using the quantum breathing mode.

\section{Acknowledgements}
We thank Hauke Thomsen for providing
the results for classical systems.
This work is supported by the Deut\-sche
Forschungs\-ge\-mein\-schaft via SFB-TR24, project A5 and a grant for 
CPU time at the HLRN.


\section*{References}

\end{document}